\documentclass[twocolumn]{aastex631}

\shorttitle{AASTeX v6.31 Sample article}
\shortauthors{Zapata et al.}
\graphicspath{{./}{figures/}}
\newcommand{\dechms}[4]{$#1^{\rm h}#2^{\rm m}#3\mbox{$^{\rm s}\mskip-7.6mu.\,$}#4$}

\newcommand{\decdms}[4]{$-#1^{\circ}#2'#3\mbox{$''\mskip-7.6mu.\,$}#4$}
\usepackage{graphicx}

\begin{document}

\title{Magnetic Fields in Massive Star-forming Regions (MagMaR) IV: Tracing the Magnetic Fields in the O-type protostellar system IRAS 16547$-$4247}

\correspondingauthor{Luis A. Zapata}

\email{l.zapata@irya.unam.mx}

\author[0000-0003-2343-7937]{Luis A. Zapata}
\affil{Instituto de Radioastronom\'\i a y Astrof\'\i sica, Universidad Nacional Aut\'onoma de M\'exico, P.O. Box 3-72, 58090, Morelia, Michoac\'an, M\'exico}

\author[0000-0001-5811-0454]{Manuel Fern\'andez-L\'opez}
\affiliation{Instituto Argentino de Radioastronom\'\i a (CCT-La Plata, CONICET; CICPBA), C.C. No. 5, 1894, Villa Elisa, Buenos Aires, Argentina}

\author[0000-0002-7125-7685]{Patricio Sanhueza}
\affil{National Astronomical Observatory of Japan, National Institutes of Natural Sciences, 2-21-1 Osawa, Mitaka, Tokyo 181-8588, Japan}
\affil{ Department of Astronomical Science, SOKENDAI (The Graduate University for Advanced Studies), 2-21-1 Osawa, Mitaka, Tokyo 181-8588, Japan}

\author[0000-0002-3829-5591]{Josep M.,Girart}
\affil{Institut d'Estudis Espacials de Catalunya (IEEC), c/Gran Capita, 2-4, E-08034 Barcelona, Catalonia, Spain}

\author[0000-0003-2343-7937]{Luis F. Rodr\'\i guez}
\affil{Instituto de Radioastronom\'\i a y Astrof\'\i sica, Universidad Nacional Aut\'onoma de M\'exico, P.O. Box 3-72, 58090, Morelia, Michoac\'an, M\'exico}

\author[0000-0002-3583-780X]{Paulo Cortés}
\affil{Joint ALMA Observatory, Alonso de Córdova 3107, Vitacura, Santiago, Chile}
\affil{National Radio Astronomy Observatory, 520 Edgemont Road, Charlottesville, VA 22903, USA}

\author[0000-0003-2777-5861]{Koch, Patrick}
\affil{Institute of Astronomy and Astrophysics, Academia Sinica, No.1, Sec. 4, Roosevelt Road, Taipei 10617, Taiwan}

\author[0000-0003-3315-5626]{Maria T.\ Beltr\'an}
\affiliation{INAF-Osservatorio Astrofisico di Arcetri, Largo E. Fermi 5, I-50125 Firenze, Italy}

\author[0000-0002-8557-3582]{Kate Pattle}
\affiliation{Department of Physics and Astronomy, University College London, Gower Street, London WC1E 6BT, UK}

\author[0000-0002-1700-090X]{Henrik Beuther}
\affiliation{Max Planck Institute for Astronomy, Konigstuhl 17, D-69117 Heidelberg, Germany}

\author[0000-0002-0028-1354]{Piyali Saha}
\affiliation{National Astronomical Observatory of Japan, National Institutes of Natural Sciences, 2-21-1 Osawa, Mitaka, Tokyo 181-8588, Japan}

\author[0000-0001-9822-7817]{Wenyu Jiao}
\affiliation{Department of Astronomy, School of Physics, Peking University, Beijing, 100871, People's Republic of China}
\affiliation{Kavli Institute for Astronomy and Astrophysics, Peking University, Haidian District, Beijing 100871, People’s Republic of China}

\author[0000-0001-5950-1932]{Fengwei Xu}
\affiliation{I. Physikalisches Institut, Universität zu Köln, Zülpicher Str. 77, D-50937 Köln, Germany}
\affiliation{Kavli Institute for Astronomy and Astrophysics, Peking University, Beijing 100871, People's Republic of China}
\affiliation{Department of Astronomy, School of Physics, Peking University, Beijing, 100871, People's Republic of China}

\author[0000-0003-2619-9305]{Xing Walker Lu}
\affiliation{Shanghai Astronomical Observatory, Chinese Academy of Sciences 80 Nandan Road, Xuhui, Shanghai 200030, China}

\author[0000-0001-5950-1932]{Fernando, Olguin}
\affiliation{Institute of Astronomy and Department of Physics, National Tsing Hua University, Hsinchu 30013, Taiwan; folguin@phys.nthu.edu.tw}

\author[0000-0003-1275-5251]{Shanghuo Li}
\affiliation{Max Planck Institute for Astronomy, Konigstuhl 17, D-69117 Heidelberg, Germany}

\author[0000-0003-3017-4418]{Ian W. Stephens} 
\affiliation{Department of Earth, Environment, and Physics, Worcester State University, Worcester, MA 01602, USA}

\author[0000-0001-7379-6263]{Ji-hyun Kang}
\affiliation{Korea Astronomy and Space Science Institute, 776 Daedeok-daero, Yuseong, Daejeon 34055, Republic of Korea}

\author[0000-0002-8691-4588]{Yu Cheng}
\affiliation{National Astronomical Observatory of Japan, 2-21-1 Osawa, Mitaka, Tokyo 181-8588, Japan}

\author[0000-0002-7497-2713]{Spandan Choudhury}
\affiliation{Korea Astronomy and Space Science Institute, 776 Daedeok-daero Yuseong-gu, Daejeon 34055, Republic of Korea}

\author[0000-0002-6752-6061]{Kaho Morii}
\affiliation{Department of Astronomy, Graduate School of Science, The University of Tokyo, 7-3-1 Hongo, Bunkyo-ku, Tokyo 113-0033, Japan}
\affiliation{National Astronomical Observatory of Japan, National Institutes of Natural Sciences, 2-21-1 Osawa, Mitaka, Tokyo 181-8588, Japan}

\author[0000-0003-0014-1527]{Eun Jung Chung}
\affiliation{Korea Astronomy and Space Science Institute, 776 Daedeokdae-ro, Yuseong-gu, Daejeon, Republic of Korea}

\author[0000-0002-6668-974X]{Jia-Wei Wang}
\affiliation{Institute of Astronomy and Department of Physics, National Tsing Hua University, Hsinchu 30013, Taiwan}
 
\author[0000-0001-7866-2686]{Jihye Hwang}
\affil{Korea Astronomy and Space Science Institute (KASI), 776 Daedeokdae-ro, Yuseong-gu, Daejeon 34055, Republic of Korea}

\author[0000-0002-9907-8427]{A-Ran Lyo}
\affiliation{Korea Astronomy and Space Science Institute, 776 Daedeokdae-ro, Yuseong-gu, Daejeon 34055, Republic of Korea}

\author[0000-0003-2384-6589]{Qizhou Zhang} 
 \affiliation{Center for Astrophysics, Harvard \& Smithsonian, 60 Garden Street, Cambridge, MA 02138, USA}
 
 \author[0000-0002-9774-1846]{Huei-Ru Vivien Chen}
\affiliation{Institute of Astronomy and Department of Physics, National Tsing Hua University, Hsinchu 300044, Taiwan}
 


\begin{abstract}
The formation of the massive stars, and in particular, the role that the magnetic fields play in their early evolutionary phase is still far 
from being completely understood. Here, we present Atacama Large Millimeter/Submillimeter Array (ALMA) 1.2 mm full polarized 
continuum, and H$^{13}$CO$^+$(3$-$2), CS(5$-$4), and HN$^{13}$C(3$-$2) line observations with a high angular resolution 
($\sim$0.4$''$ or 1100 au). In the 1.2 mm continuum emission, we reveal a dusty envelope surrounding the massive protostars, IRAS16547-E and IRAS16547-W, 
with dimensions of $\sim$10,000 au. This envelope has a bi-conical structure likely carved by the powerful thermal radio jet present in region. 
The magnetic fields vectors follow very-well the bi-conical envelope. The polarization fraction is $\sim$2.0\% in this region.  
Some of these vectors seem to converge to IRAS 16547-E, and IRAS 16547-W, the most massive protostars. Moreover, the velocity 
fields revealed from the spectral lines H$^{13}$CO$^+$(3$-$2), and HN$^{13}$C(3$-$2) show velocity gradients with a 
good correspondence with the magnetic fields, that maybe are tracing the cavities of molecular outflows or maybe in some parts infall. 
We derived a magnetic field strength in some filamentary regions that goes from 2 to 6.1\,mG.  We also find that the CS(5$-$4) molecular 
line emission reveals multiple outflow cavities or bow-shocks with different orientations, some of which seem to follow the NW-SE radio thermal jet.
\end{abstract}


\section{introduction} \label{sec:intro}

One of the key ingredients influencing the formation of the massive stars (M$_*$ $>$ 10 M$_\odot$) is very likely the magnetic field.  
The magnetic fields can regulate or even prevent the infall of material toward the nascent massive stars \citep{Hull2019}.  At the moment, 
there is increasing evidence, for example, that the magnetic fields drive the energetic outflows and jets that emanate from protostars, 
which remove angular momentum from the system, thereby allowing the material to be transported directly onto the central protostar \citep{bla1982,chin2017,dev2020,lop2023,lau2023}. 
However, the magnetic fields in most of the massive star forming regions are not well-understood due to the large distances of these regions, and their scarcity. 
Recently, thanks to the new sensitive and superb angular resolution observations of the thermal dust polarization (using {\it e.g.}  
Atacama Large Millimeter/Submillimeter Array (ALMA), The Jansky Very Large Array (JVLA) or The Submillimeter Array (SMA),
there has been a surge of new cases where the morphology and the strength are starting to be known with a fascinating clarity 
\citep{liu2023,par2023,koch2022,cor2021,San2021,man2021,pau2021,Beu2020,Liu2020,Bel2019}.  

For example, in the cases of the massive star forming regions IRAS 18089$-$1732,  G327.3, and W51 (in several cores), it has been proposed that the magnetic 
fields are tracing spiral-like accreting structures mapped in the 1.2 mm continuum emission arising from thermal dust, and the molecular line 
emission \citep{koch2022,San2021,Beu2020}. In all cases, the magnetic field morphology suggests that the angular momentum is high enough to 
twist the magnetic field lines, and the material can infall to the forming massive protostars.  

Embedded in a massive dusty core with an estimated mass of 1.3 $\times$ 10$^3$ M$_\odot$ is found the luminous (with a bolometric 
luminosity of 6.2 $\times$ 10$^4$ L$_\odot$) infrared source IRAS16547$-$4247  \citep{gar2003}.  This massive source is located at a 
distance of 2.9 $\pm$ 0.6 kpc \citep{rod2008}, and is powering one of the radio thermal jets related to a star forming region \citep[see also the case of DG Tau;][]{rod2012}.  
The radio thermal jet has a NW-SE orientation \citep{rod2008,fra2009}, and is part of an energetic molecular outflow traced by the CH$_3$OH, CO, 
and H$_2$ thermal emission \citep{gar2007,bro2003, Hig2015}.  In the middle of IRAS16547$-$4247 there are two massive, and compact ($\sim$500 au) 
rotating circumstellar disks reported recently, IRAS16547-Ea, and IRAS16547-Eb \citep{tan2020, zap2019}. One of these disks, IRAS16547-Ea, is energizing the ionized 
free-free thermal jet, and the molecular outflow \citep{tan2020}. From the kinematics of the circumbinary disk that surrounds 
both compact disks \citep{zap2019},  the estimated mass for the central objects (the binary system)  is around 25 $\pm$ 3 M$_\odot$, 
resulting in true O-type protostars \citep{zap2019}.            
 
In this study, we present 1.2 mm dust polarization, and molecular line ALMA observations 
of the rotating, and massive hot molecular core embedded in the high-mass star-forming region IRAS16547$-$4247.  These observations reveal 
for the first time the detail structure of the magnetic fields associated with the dust, and line emission from the envelope 
surrounding to  IRAS16547$-$4247 (throughout our study we also refer to this source as IRAS16547).  

\begin{figure*}[ht!]
\epsscale{1.2}
\plotone{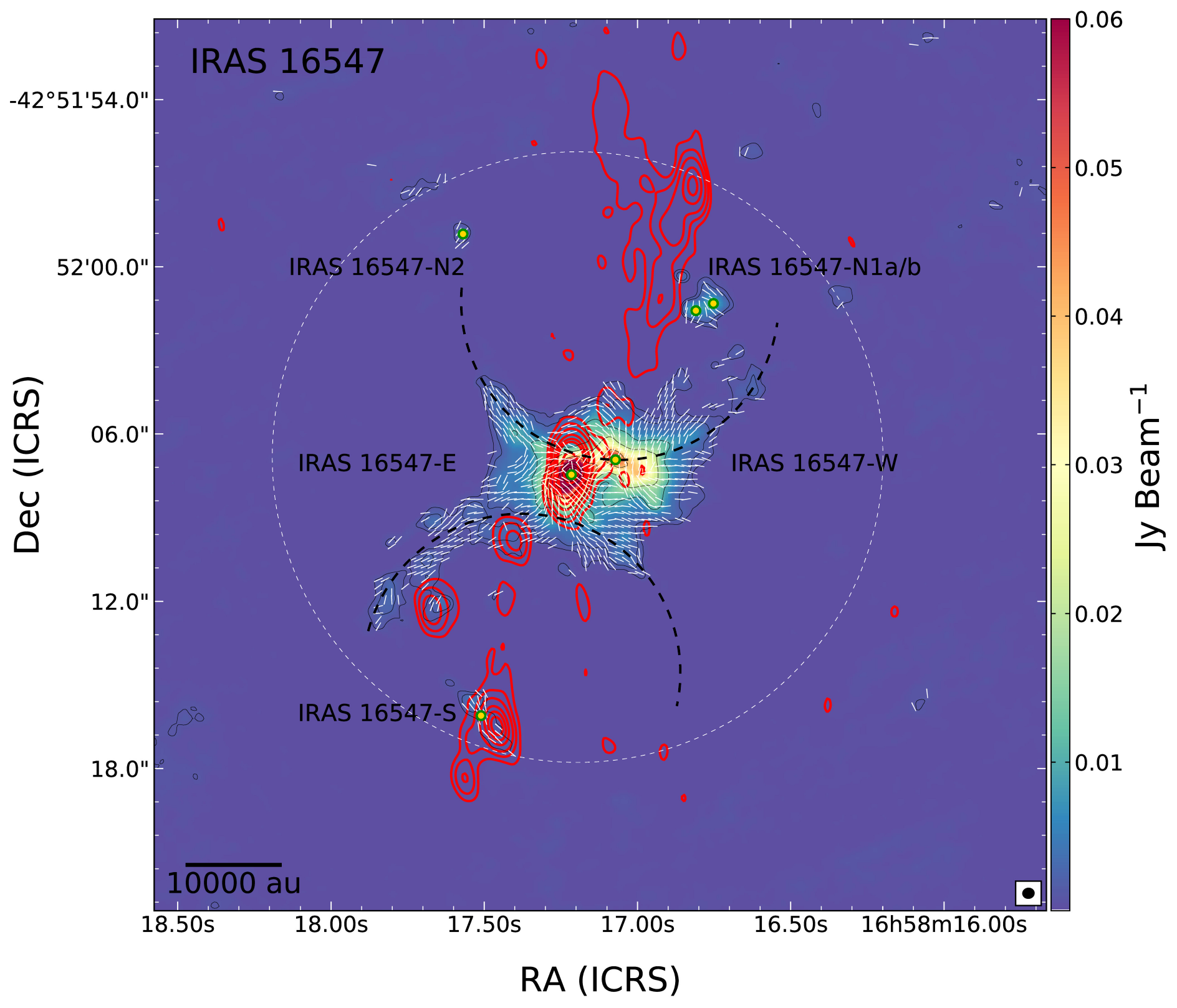}
\caption{\scriptsize ALMA magnetic field vectors overlaid onto the resulting 1.2 mm continuum emission (Stokes I) from IRAS 16547$-$4247. 
The white vector segments trace the polarization percentage.  The vector segment is displayed more or less per every beam, 
using only data with a 2$\sigma$ cut in the debiasing process. The black contours are from $-$0.5, 0.5, 1, 5, 10, 20, 40, and 60 times 0.2 mJy Beam$^{-1}$ the RMS-noise of the image.  
The red contours are from $-$5, 5, 10, 20, 40, 60, 80, and 180 times the RMS-noise of the 
VLA 3.6 cm continuum emission, which is 20 $\mu$Jy Beam$^{-1}$.   The 3.6 cm continuum emission delineates the radio thermal jet  with  a NW-SE 
orientation reported by \citet{gar2003,rod2008}. The synthesized beam (half-power contour) of the ALMA continuum image is shown in the bottom-right corner. 
The scale bar in the right represent the peak flux of the ALMA 1.2 mm continuum emission. The dashed lines trace more or less the bi-conical dusty envelope revealed in this study.
The white circle traces the primary beam or the FWHM of 22.7$''$. The yellow dots trace the positions of the sources reported in Table  ~\ref{table1}.}  
\label{fig:fig1}
\end{figure*}

\begin{table*}[!]
\caption{Physical parameters of the 1.2 mm continuum sources}
\resizebox{\textwidth}{!}{
\begin{tabular}{l*{8}{c}rr}
\multicolumn{10}{c}{}&\\
\hline
\hline
\multicolumn{2}{c}{Source}&
\multicolumn{2}{c}{Position}&
\multicolumn{3}{c}{Deconvolved Size}&
\multicolumn{2}{c}{Flux}&
\multicolumn{1}{c}{Mass}&\\
\hline
&{} & $\alpha_{2000}$ &{$\delta_{2000}$} & {$\theta_{maj}$} &{$\theta_{min}$} &P.A.&{Peak} &{Integrated}& \\
&{} & [h m s] & [$^{\circ }$ $^{\prime }$ $^{\prime\prime }$ ] & [mas] & [mas] &[$^{\circ }$] & {[mJy Beam$^{-1}$]} &{[mJy]}&[M$_\odot$]\\ 
\hline
\hline
IRAS16547N1a  & & 16 58 16.765 $\pm$ 0.02$''$   &  $-$42 52 01.37 $\pm$ 0.01$''$   & 830    $\pm$ 70     &    550  $\pm$ 50    &116 $\pm$ 10  &  9.5 $\pm$ 0.5  & 34.5 $\pm$ 2    & 1.4 (50 K)\\  
IRAS16547N1b  & & 16 58 16.803 $\pm$ 0.02$''$   &  $-$42 52 01.42 $\pm$ 0.01$''$   & 1013  $\pm$ 60     &    628  $\pm$ 40    &  96 $\pm$ 5    &12.3 $\pm$ 0.6  & 57.5 $\pm$ 3    & 2.2 (50 K)\\
IRAS16547N2    & & 16 58 17.593 $\pm$ 0.05$''$   &  $-$42 51 58.89 $\pm$ 0.05$''$   & 1540  $\pm$ 140   &  1390  $\pm$ 130  &  97 $\pm$ 40  &  2.0 $\pm$ 0.2  & 29.0  $\pm$ 3   &1.1  (50 K) \\ 
IRAS16547E      & & 16 58 17.217 $\pm$ 0.02$''$   &  $-$42 52 07.46 $\pm$ 0.02$''$   & 862    $\pm$ 65     &    532  $\pm$ 47    & 174 $\pm$ 10 & 174 $\pm$ 9.0  & 637 $\pm$ 42   &6.1  (150 K)\\
IRAS16547W     & & 16 58 17.062 $\pm$ 0.10$''$   &  $-$42 52 07.03 $\pm$ 0.04$''$   & 2490  $\pm$ 250   &  1000  $\pm$ 100  &   77 $\pm$ 4   & 35.0 $\pm$ 3     & 535 $\pm$ 52   &5.1  (150 K)  \\
IRAS16547S      & & 16 58 17.535 $\pm$ 0.05$''$   &  $-$42 52 15.64 $\pm$ 0.04$''$   & 958    $\pm$ 153   &    533  $\pm$ 104  &   53 $\pm$ 10 &   4.0 $\pm$ 0.5  &16.0 $\pm$ 2     &0.23 (50 K) \\   
\hline
\hline
\end{tabular}}
\tablecomments{These physical parameters were obtained with the CASA IMFIT task. In the mass estimation, we have also included which dust temperature was used for a such purpose.}
\label{table1}
\end{table*}

\section{Observations}

ALMA observations to IRAS16547$-$4247 were taken on 2018 December under the program 2017.1.00101.S with PI: Sanhueza, P.  At that time, 
the array included 45 antennas with diameters of only 12 m, and covering baselines from 15 to 783 m (12 to 652 k$\lambda$).  
The observations were pointed at the phase center  $\alpha_{J2000.0}$ = \dechms{16}{58}{17}{11}, and
$\delta_{J2000.0}$ = \decdms{42}{52}{07}{06}. The thermal dust, and molecular gas emission in the Stokes I, Q, U, and V from IRAS16547$-$4247 
were covered with a single pointing with a Full Width Half Maximum (FWHM) of 22.7$''$. Henceforth, we only consider the continuum sources and the polarization vectors inside of the FWHM as real. 
The Largest Angular Scale (LAS) that is recovered by these observations is 6.3$''$.  The total on-source integration time was 45 min.

The weather conditions during the observations were considerably well with an average precipitable water vapor of about 1.9 mm. 
The phase Root Mean Square (RMS) was only 19.8$^\circ$. The atmospheric phase oscillation was reduced by the simultaneously 
observation of the 183 GHz water line with water vapor radiometers. Quasars J1751$+$0939, J1517$-$2422, and J1733$-$3722 were used to perform 
the bandpass calibration, the gain/atmospheric fluctuations, and the flux amplitude calibration. Some of the quasar scans were repeated for different calibrations.

The ALMA digital correlator was configured with five Spectral Windows (SPWs) intended to detected different molecular species in these 
millimeter wave bands, and the adjacent continuum.  Three SPWs were centered at  rest frequencies of 243.518, 245.518, and 257.521 GHz, 
and have a bandwidth of 1.875 GHz with a number of channels 1920, which gives a channel spacing of  976.5 kHz or 1.195 km s$^{-1}$. 
In order to construct the millimeter continuum emission for IRAS16547$-$4247, we only used free-line channels in the SPWs. 
The rest of SPWs were centered to detect the H$^{13}$CO$^+$(3$-$2), and the HN$^{13}$C(3$-$2) spectral lines (at rest frequencies of 260.25554 GHz, 
and 261.26331 GHz, respectively) with a channel spacing of 0.281 km s$^{-1}$ or 243.7 kHz. We also detected the spectral line CS(5$-$4) 
at a rest frequency of 244.93555 GHz. The CS molecule is a high density and classical outflow tracer \citep{pato2010}. 
The bulk of thermal emission only from these three spectral lines are discussed in this study. 

\begin{figure*}[ht!]
\epsscale{0.53}
\plotone{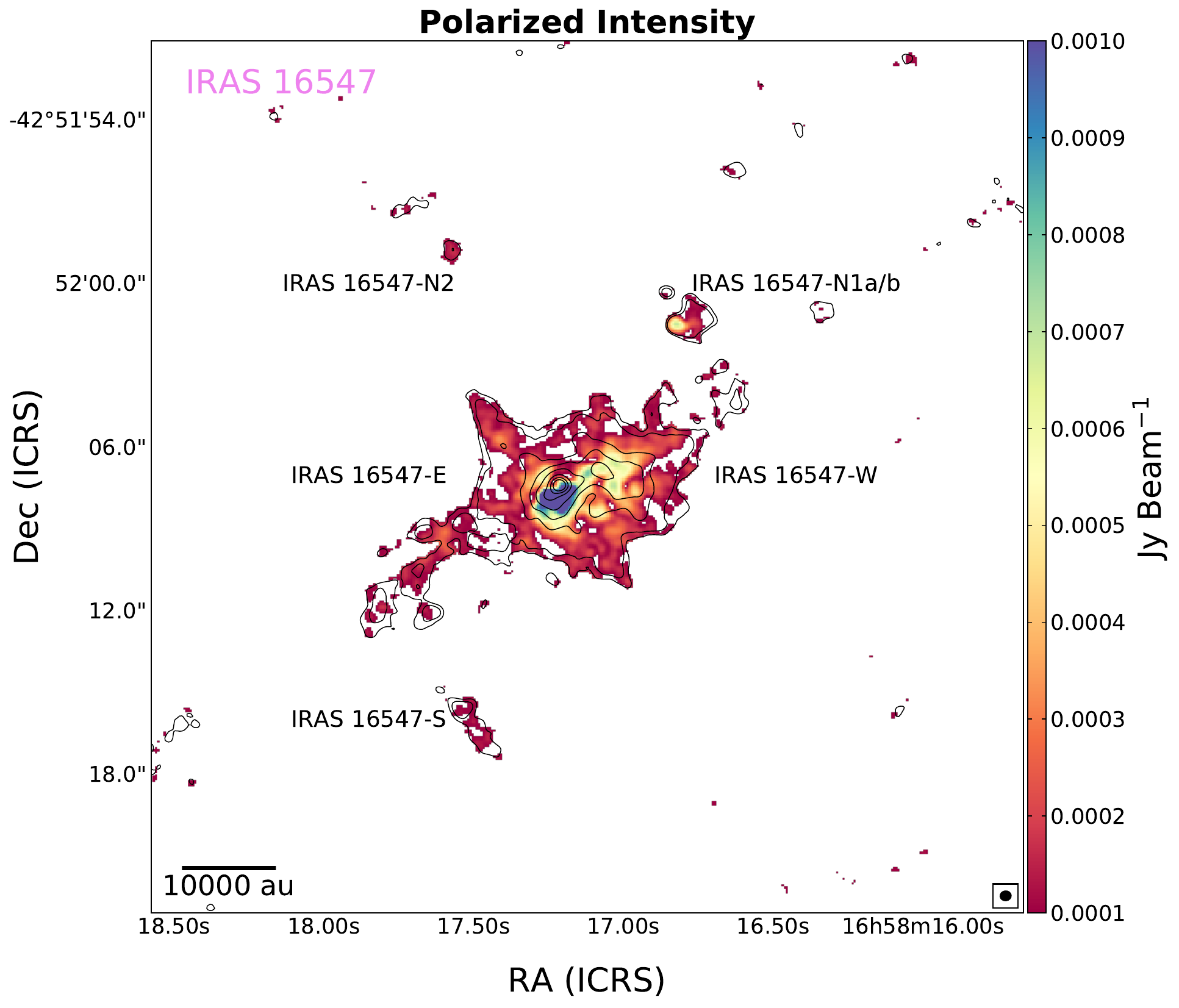}
\epsscale{0.5}
\plotone{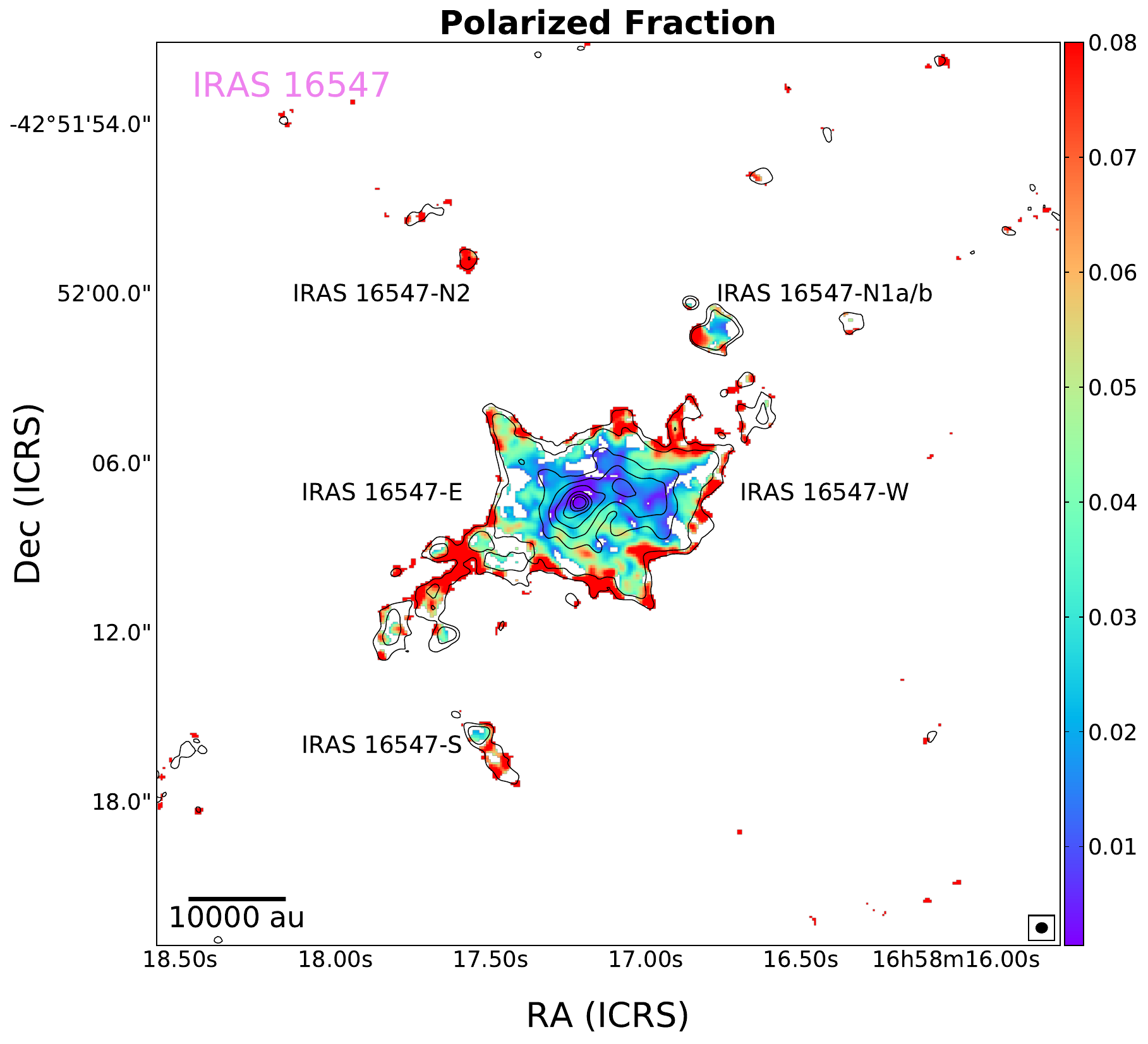}
\caption{\scriptsize Left Panel: ALMA Polarization Intensity map overlaid with the 1.2 mm continuum emission from IRAS 16547$-$4247. 
Right Panel: Same as the left panel, but showing the ALMA Polarization Fraction. In both panels, the black contours are from $-$0.5, 0.5, 1, 5, 10, 20, 40,  and 60 
times 0.2 mJy Beam$^{-1}$ the RMS-noise of the image.  The synthesized beam (half-power contour) of the ALMA continuum
 image is shown in the bottom-right corner. The scale bar in the right represents the de-biased polarization intensity, and polarization fraction.
 These maps were constructed with a threshold of 2$\sigma$ for the polarization.  } 
\label{fig:fig2}
\end{figure*}

The ALMA data were calibrated, analyzed, and imaged in a standard manner using Common Astronomy Software Applications (CASA) VERSION  5.4.0$-$68 \citep{casa2022}.  
The resulting images were Fourier transformed, deconvolved, and restored using the CASA TCLEAN task. The resulting RMS-noise for the continuum image 
of the Stokes I is 0.2 mJy Beam$^{-1}$ with a synthesized beam of 0.44$''$ $\times$ 0.39$''$ and a Position Angle (PA) of $-$83.1$^\circ$.  
As in other cases \citep[see][]{zap2023}, the RMS-noise is higher than the theoretical (which is about 0.05 mJy Beam$^{-1}$). This is  likely due to the strong mm 
continuum source localized in IRAS16547$-$4247, that did not allow us to reach the theoretical RMS noises. Contrary to the Stokes I, for the Stokes Q, U, and V
 images the RMS noise level is 0.05 mJy Beam$^{-1}$, very close to the theoretical value. We construct the images of the debiased linear polarized intensity ($\sqrt{Q^2+U^2-0.5[dQ^2+dU^2]}$), 
 the linear de-biased polarization fraction ($\sqrt{Q^2+U^2}$/I), and the electric vector position angle (0.5$\arctan$(U/Q)), and they are present in 
 Figures ~\ref{fig:fig1}  and ~\ref{fig:fig2}. The resulting RMS-noise for the line images in Stokes I is 3 mJy Beam$^{-1}$ per channel spacing with a synthesized 
 beam of 0.36$''$ $\times$ 0.32$''$ and a PA of $-$88.1$^\circ$. We obtained values similar to the ALMA theoretical RMS-noise for the spectral line images.

Phase self-calibration (using interval solutions that included:  inf., 60s and 15s) was made using the continuum (Stokes I) as a model. 
The spectral line channels were then corrected with the acquired solutions. The improvement in the continuum RMS-noise was almost a factor of 10 
for the continuum, and 3 for the line emission.

 \begin{figure}[ht!]
\epsscale{1.1}
\plotone{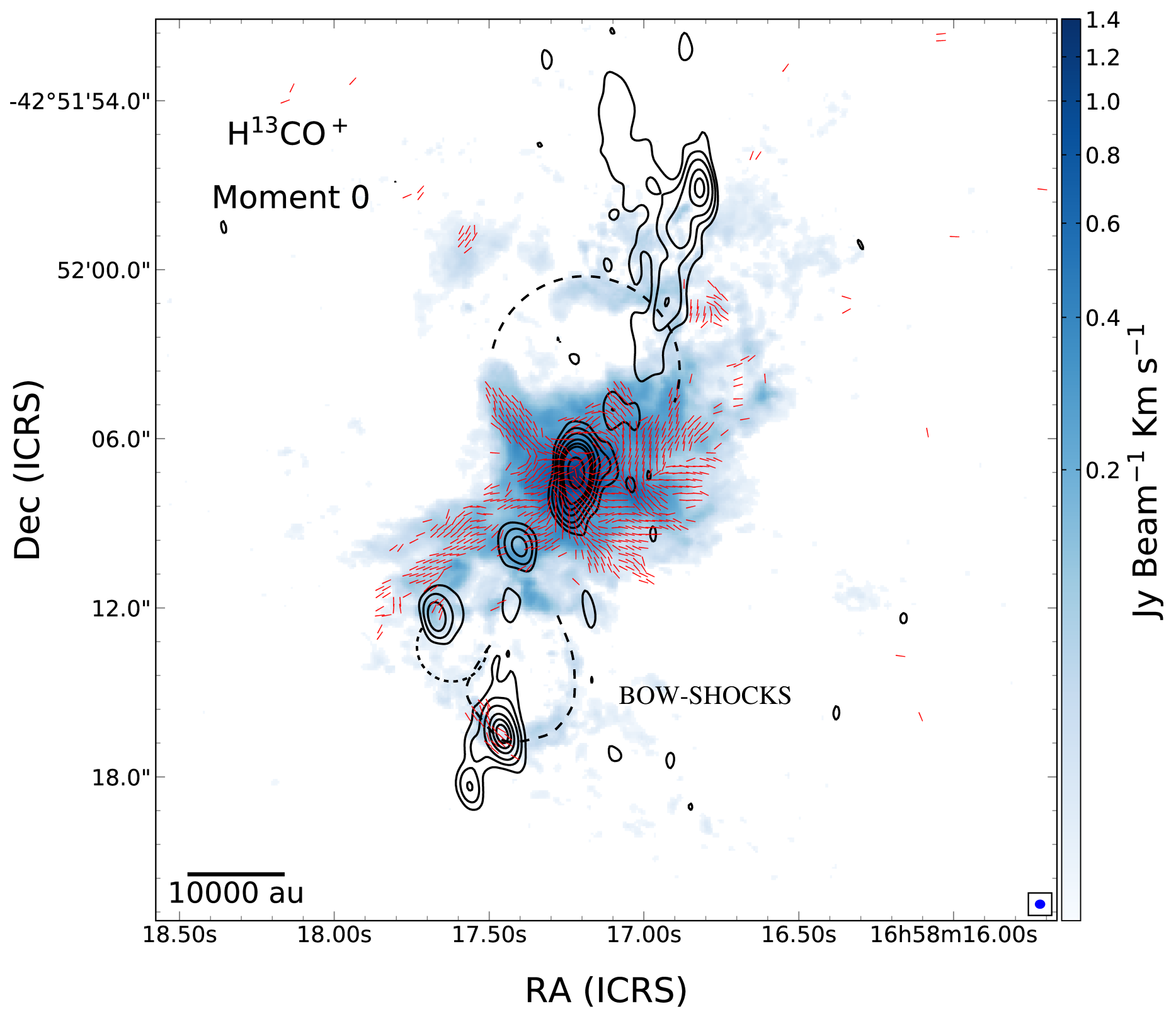}  \vspace{0.5cm}
\plotone{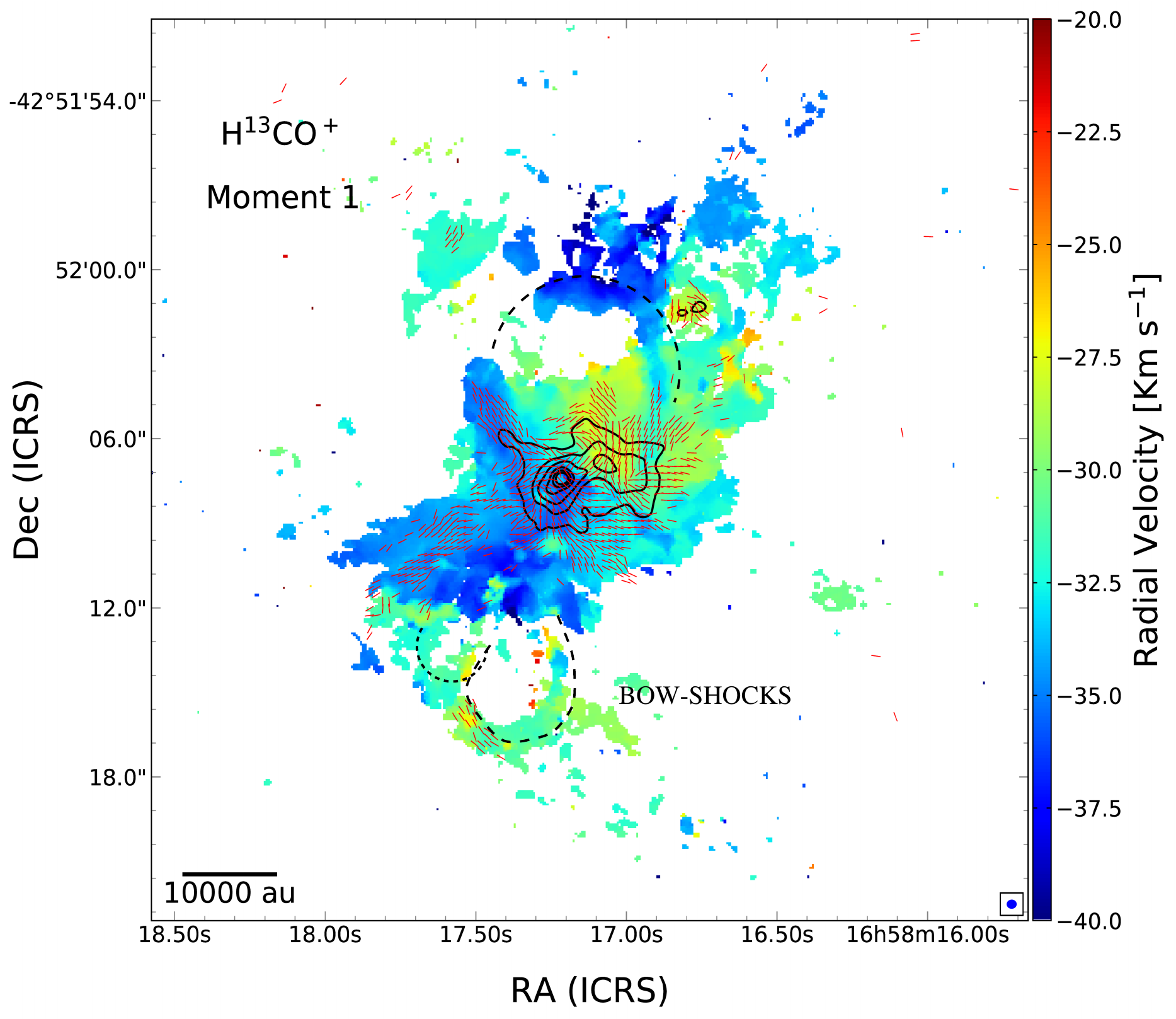}
\caption{\scriptsize ALMA magnetic field vectors overlaid onto the resulting H$^{13}$CO$^+$(3$-$2) line Moment zero (upper) and Moment one maps (lower) from IRAS 16547$-$4247. 
The red vector segments trace the polarization percentage. The vector segment is displayed more or less per every beam, 
using only data with a 2$\sigma$ cut in the debiasing process (Same as Figure \ref{fig:fig1}). In the upper panel, the black contours are 
from $-$5, 5, 10, 20, 40, 60, 80, and 180 times the RMS-noise of the 
VLA 3.6 cm continuum emission, which is 20 $\mu$Jy Beam$^{-1}$.   The 3.6 cm continuum emission delineates a radio free-free jet  with NW-SE 
orientation reported by \citet{gar2003,rod2008}. The scale bar in the right represents the peak flux of the ALMA H$^{13}$CO$^+$(3$-$2) line emission. 
In the lower panel, the black contours are from $-$0.5, 0.5, 1, 5, 10, 20, 40, and 60 times 0.2 mJy Beam$^{-1}$ the RMS-noise of the continuum image. 
The scale bar in the right represents the the weighted mean radial velocity of the ALMA H$^{13}$CO$^+$(3$-$2) line emission.
The synthesized beam (half-power contour) of the ALMA line image is shown in the bottom-right corner of each panel. The dashed curves trace the multiple bow-shocks and outflow cavities
discussed in the main text.
\label{fig:fig3}}
\end{figure}

\section{Results and Discussion}

\subsection{ALMA polarization observations of IRAS 16547$-$4247} 

In Figure~\ref{fig:fig1}, we show one of the main results of this study, the inferred structure of the magnetic fields (white vectors) plotted on 
the 1.2 mm continuum emission from IRAS16547$-$4247. In order to compute the magnetic field orientations, we are assuming 
that these structures are orthogonal to the linear polarization orientations of the Electric Vector Position Angles (EVPAs), a result from 
grains aligned by Radiative Torques to precess around the magnetic field direction \citep[B-RAT alignment;][]{hog2008}. Both, the magnetic field orientations (vectors), 
and the 1.2 mm continuum emission (black contours) in IRAS16547$-$4247 are arising from six continuum sources with physical parameters 
described in Table~\ref{table1}. The faintest source, IRAS16547N2, is at a 10$\sigma$ level.
The strongest sources of 1.2 mm continuum emission are associated with IRAS 16547-E, and IRAS 16547-W.  IRAS 16547-E is 
in addition related to a powerful radio free-free thermal jet (in red contours in Figure~\ref{fig:fig1}) with a NW-SE orientation that is emanating from one of the compact 
circumstellar disk called IRAS16547-Ea, and that surrounds a true young O-type protostar \citep{tan2020}.  
For the case of IRAS 16547-W, this source is also associated with hot molecular emission from complex organic molecules 
as revealed in Figure~\ref{fig:fig2} of \citet{zap2019}. This mm continuum source could also be related to a massive protostar \citep{fra2009,zap2019}. 
The present ALMA observations are sensitive to larger scales ($\sim$6.3$''$), and cannot separate the thermal dust, and free-free emission 
at these millimeter wavelengths arising from IRAS 16547-E, and IRAS 16547-W, see for example the observations at a wavelength of 3 mm from \citet{tan2020}, where they separate both emissions.  
However, the free-free emission related to this object, and at these wavelengths is indeed very low (3$\%$, arising from IRAS 16547-E) as discussed in \citet{zap2019}. We capture the dusty envelope 
surrounding the massive protostars. This dusty and large envelope, with spatial scales that reaches $\sim$10,000 au, has a bi-conical structure 
with the open angles that follow the same orientation as that of the radio thermal jet. The morphology of the dusty envelope could be explained in terms of that the NW-SE 
outflow is carving or cleaning up the dust and gas in these northern/southern orientations.  The rest of the millimeter continuum sources presented in Table~\ref{table1} 
could also be part of this bi-conical envelope, although we may be missing some more extended  continuum millimeter emission that connects all these structures.      

At this millimeter wavelength regimen, we can estimate the mass of the sources detected in this work, assuming that the emission is optically thin, 
and isothermal, and following the expression \citep{hilde1983}: 
$$
M_d=\frac{D^2S_\nu}{\kappa_\nu B_\nu (T_d)},
$$
where D is the distance to IRAS16547$-$4247, 2.9 $\pm$ 0.6 kpc \citep{rod2008}, S$_\nu$ is the flux density,  $\kappa_\nu$ = 0.0103 cm$^2$ g$^{-1}$ 
is the dust mass opacity for a gas-to-dust ratio of 100 appropriate for 1.2 mm \citep{hop1994}, and $B_\nu (T_d)$ is the Planck function for dust temperature 
T$_d$. We considered the Rayleigh$-$Jeans regime, given the observing frequency, 250 GHz. The estimated values for the mass are given in Table~\ref{table1}.
 Here, we are assuming a dust temperature T$_d$ = 150 K for IRAS 16547-E and IRAS 16547-W, given that they are associated with hot molecular 
 gas  \citep{zap2019, tan2020}, but for the rest sources we assume a T$_d$ = 50 K (this a good approximations as for example IRAS 16547-W is associated with a massive protostar).  
 The mass values for IRAS 16547-E and IRAS 16547-W are in very 
 good agreement with those values obtained by \citet{zap2015}.  Given the uncertainty in the dust temperature and the opacity, the uncertainty in the mass estimation 
 could be very large, a factor of up to 3 or 4. Using two ellipsoidal Gaussian fitting to the central bi-conical envelope (Figure~\ref{fig:fig1}), 
 we obtain a flux density of 2.2$\pm$0.2 Jy and a peak flux of 201$\pm$18 mJy Beam$^{-1}$, which corresponds to a total mass of 220 M$_\odot$, assuming a 
dust temperature of 16 K, a temperature obtained from HN$^{13}$C spectral line (see next Section). Following \citet{vic2014}, we estimate a column density of 4.0$\times$10$^{26}$ cm$^{-2}$, 
and a volumetric density of 2.2$\times$10$^{8}$ cm$^{-3}$.    

The magnetic fields follow the structure of the continuum emission very well. Some of these vectors seem to converge to IRAS 16547-E and IRAS 16547-W, the massive protostars (see Figure~\ref{fig:fig1}).    
 A possible explanation is that we are seeing a dusty shell, and that the magnetic fields mapped here are part of the envelope opened by the radio thermal jet.    
 In-falling dusty envelopes have also been proposed to explain the orientation of the magnetic fields (see, IRAS 18089$-$1732 \citep{San2021} and W51 \citep{koch2022}), 
 however, here we only leave this as a speculative possibility.

The polarized intensity, and polarization fraction from IRAS16547$-$4247 are presented in Figure~\ref{fig:fig2}. The polarized fraction shows more or less a constant intensity towards the edges of the 
dusty bi-conical envelope with a sustancial increase close to IRAS16547-E, on its southern side.  On the other hand, the polarized fraction shows a structure contrary to the
the polarized intensity with a depression of the intensity in the central part, and the brightest intensity in the edges of the dusty envelope, indicating that the mayor polarized 
radiation is coming from the most densest parts of the dusty envelope.  

\begin{figure}[ht!]
\epsscale{1.1}
\plotone{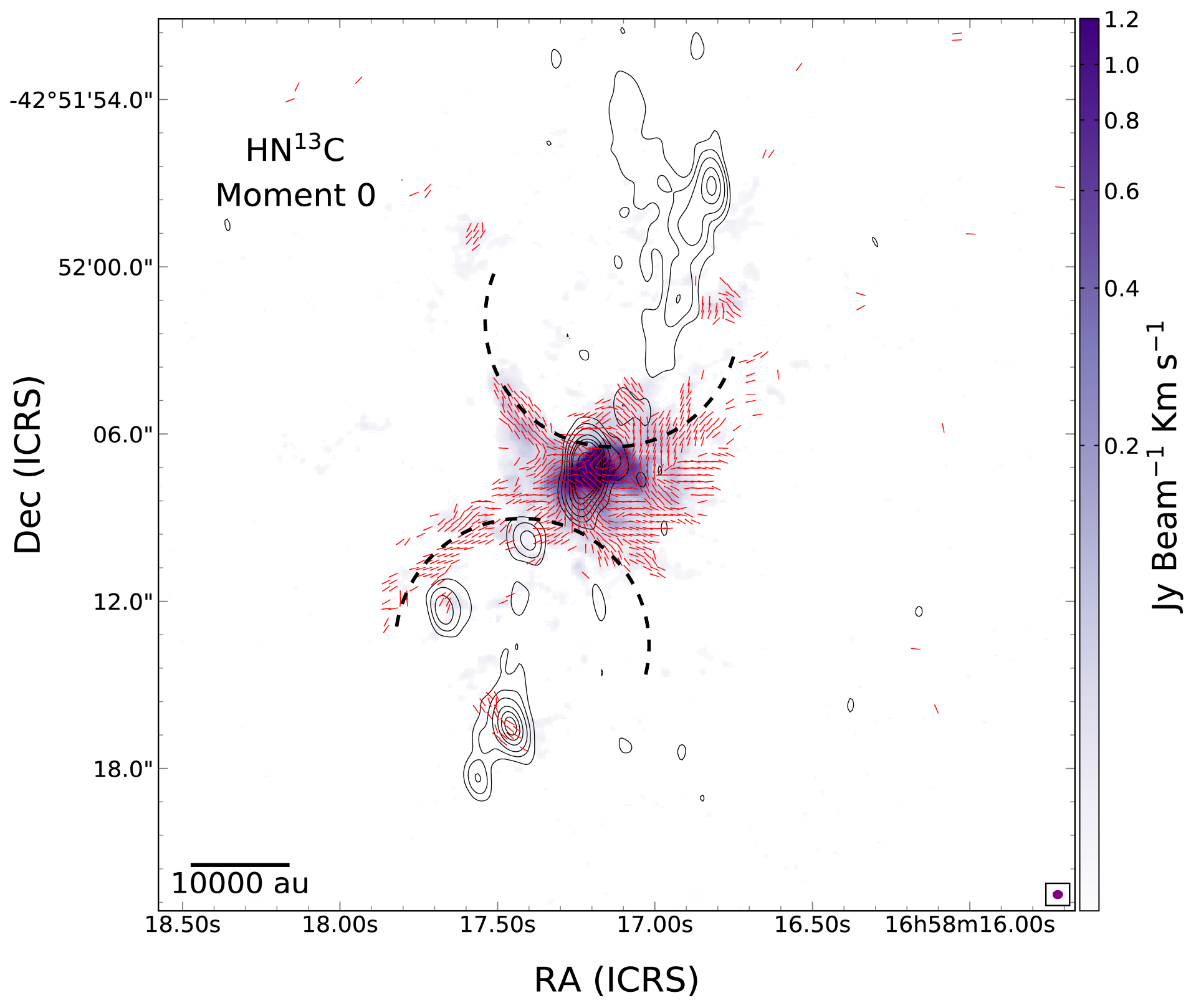}  \vspace{0.5cm}
\plotone{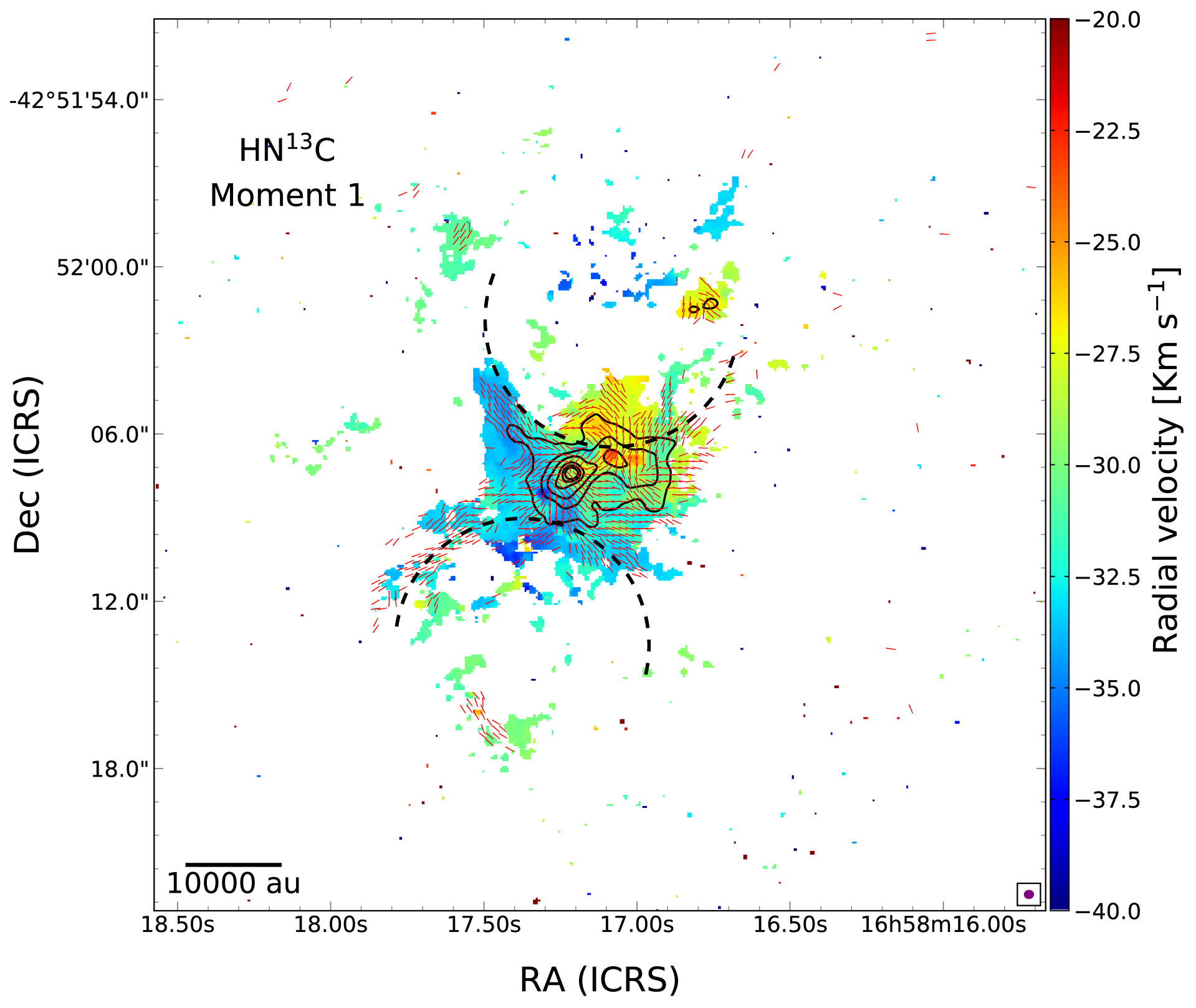}
\caption{\scriptsize ALMA magnetic field  overlaid onto the resulting HN$^{13}$C line Moment 0 (upper) and Moment 1 maps (lower) from IRAS 16547$-$4247. 
The red vector segments trace the polarization percentage. The vector segment is displayed more or less per every beam, 
using only data with a 2$\sigma$ cut in the debiasing process (Same as Figure \ref{fig:fig1}). In the upper panel, the black contours are from $-$5, 5, 10, 20, 40, 60, 80, and 180 
times the RMS-noise of the VLA 3.6 cm continuum emission, which is 20 $\mu$Jy Beam$^{-1}$.   The 3.6 cm continuum emission delineates a radio free-free jet  with NW-SE 
orientation reported by \citet{gar2003,rod2008}. The scale bar in the right represents the peak flux of the ALMA HN$^{13}$C line emission. In the lower panel, the black contours 
are from $-$0.5, 0.5, 1, 5, 10, 20, 40, 60, and 80 times 0.2 mJy Beam$^{-1}$ of the ALMA 1.2 mm continuum emission. 
The scale bar in the right represents the radial velocity of the ALMA HN$^{13}$C line emission.
The synthesized beam (half-power contour) of the ALMA line image is shown in the bottom-right corner of each panel.
The dashed lines trace more or less the bi-conical dusty envelope revealed in this study.
\label{fig:fig4}}
\end{figure}

 \begin{figure}[ht!]
\epsscale{1.1}
\vspace{-1cm}
\plotone{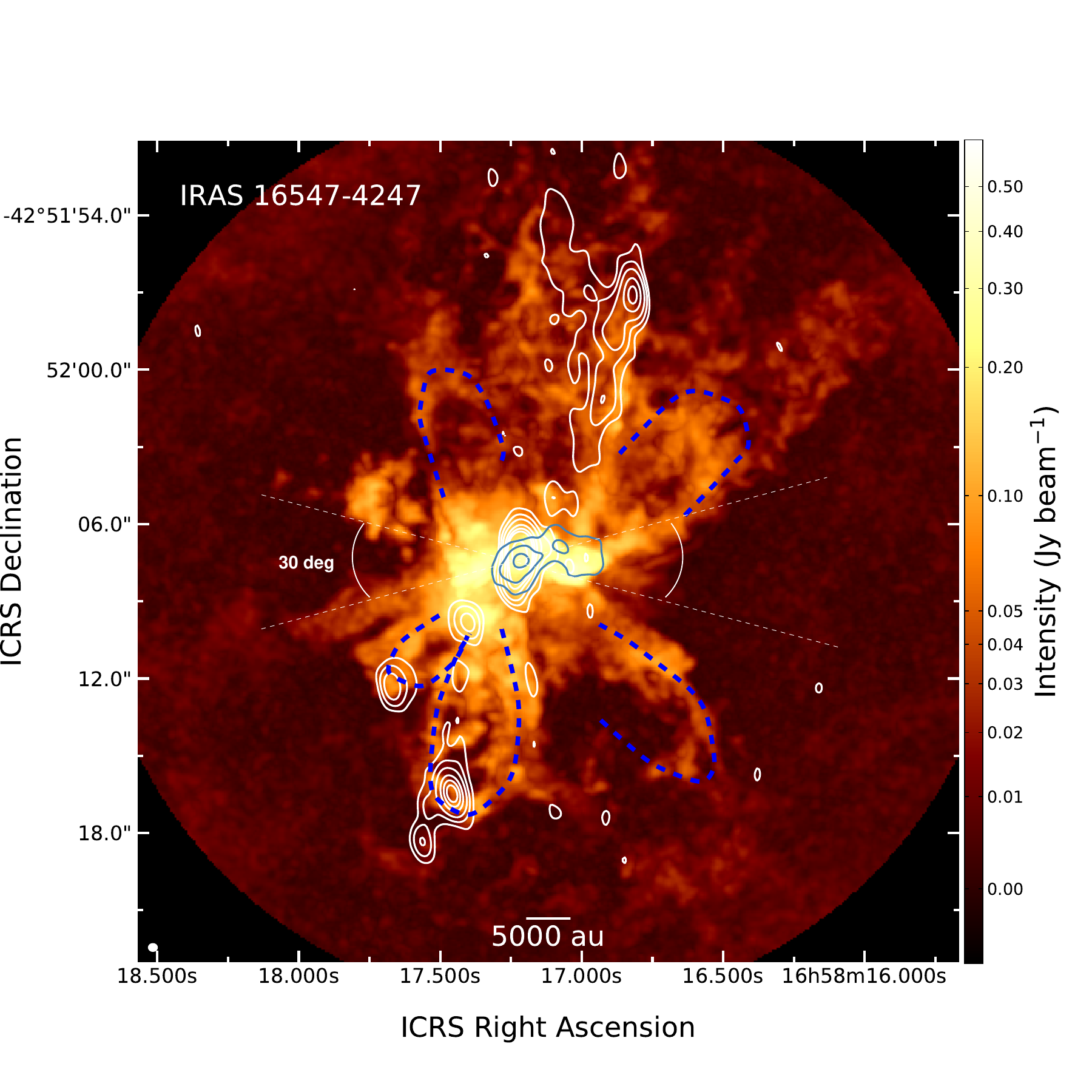}
\vspace{-1cm}
\plotone{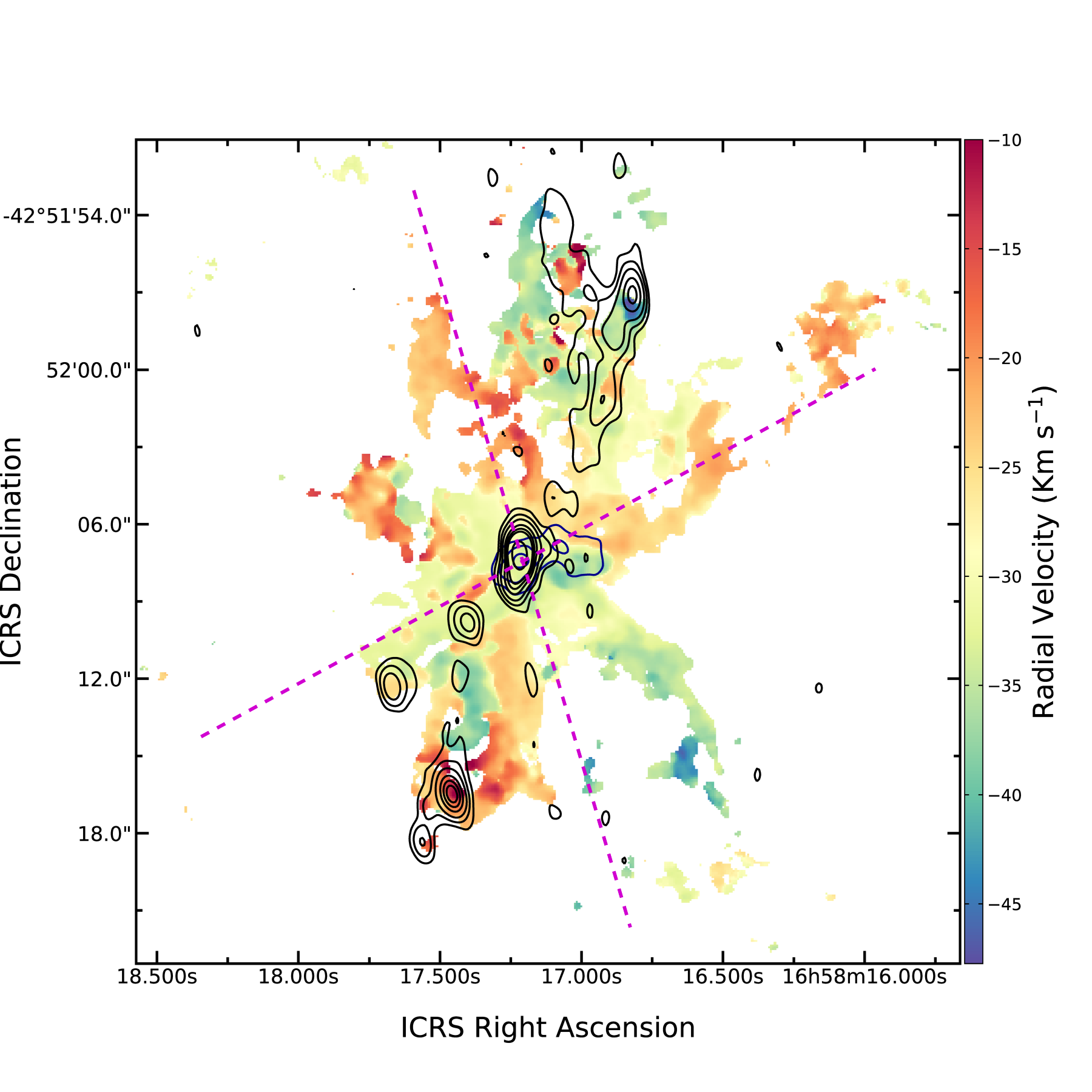}
\caption{\scriptsize Upper: ALMA moment 8 map of the CS(5$-$4) from IRAS 16547$-$4247 overlaid onto the resulting 1.2 mm continuum emission (grey contours), 
and the VLA 3.6 cm continuum emission (white contours). The grey contours are from 50, 125, and 375 times 0.2 mJy Beam$^{-1}$.  
The white contours are from $-$5, 5, 10, 20, 40, 60, 80, and 180 times the RMS-noise of the VLA 3.6 cm continuum emission, which is 20 $\mu$Jy Beam$^{-1}$.   
The 3.6 cm continuum emission delineates a radio free-free jet  with NW-SE  orientation reported by \citet{gar2003,rod2008}. 
The synthesized beam (half-power contour) of the ALMA continuum image is shown in the bottom-right corner. 
The scale bar in the right represents the peak flux of the ALMA CS line emission. The blue dashed curves trace more or less the multiple bow-shocks and outflow cavities revealed by this molecule.
 The white lines show that aperture angle of the outflow is more than 150$^\circ$ in the north and south sides. 
 Lower: Same as the upper image but with the ALMA moment 1 map of the CS(5$-$4). The scale bar in the right represents the radial velocity of the ALMA CS line emission. 
  The magenta dashed lines mark the orientation of the CO(3$-$2) ouflows reported by \citet{Hig2015}.  
\label{fig:fig5}}
\end{figure}

\begin{figure*}[ht!]
\epsscale{1.18}
\plotone{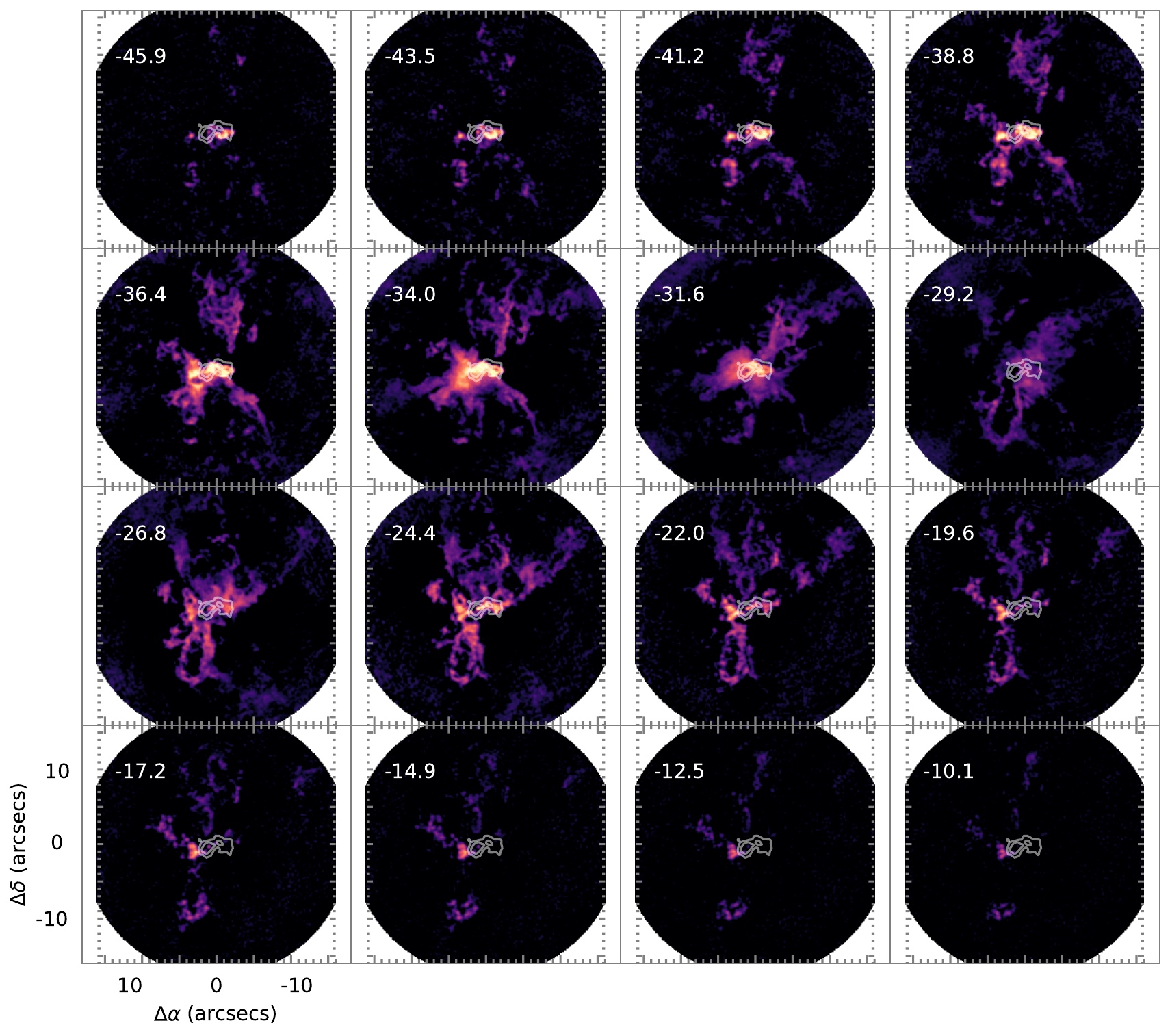}
\caption{\scriptsize ALMA velocity channel map of the CS(5$-$4) from IRAS 16547$-$4247 overlaid onto the resulting 1.2 mm continuum emission (white contours). 
The white contours are from 50, 125, and 375 times 0.2 mJy Beam$^{-1}$ the RMS-noise of the image.  The synthesized beam (half-power contour) of the ALMA continuum image is 0.36$''$ $\times$ 0.32$''$ 
and a PA of $-$88.1$^\circ$. The radial velocity is indicated in the upper left side of each channel in km s$^{-1}$. 
\label{fig:fig6}}
\end{figure*}
 
The Moment 0 (integrated intensity) and Moment 1 (intensity-weighted velocity image) maps of the H$^{13}$CO$^+$(3$-$2) emission overlaid with the 
polarization vectors that are tracing the magnetic fields in IRAS16547$-$4247, the continuum emission at 1.2 mm, and the thermal free-free jet is presented in Figure~\ref{fig:fig3}. To compute these moment maps 
we integrate over radial velocities from $-$40.9 km s$^{-1}$ to $-$15.9 km s$^{-1}$. The systemic cloud velocity of IRAS16547$-$4247 is about $-$30 km s$^{-1}$ \citep{tan2020}.  These moments show 
us that the molecular thermal emission is again tracing an envelope with bi-conical outflow cavities, but with some clear bow-shocks and outflow cavities far from the dusty envelope, 
and that coincide well with the north-south free-free thermal jet.   These bow-shocks structures or outflow cavities are also seen in the millimeter wavelength CH$_3$OH emission maps from \citet{Hig2015}, 
and are interpreted as part of the outflow emanating from here.  From the southern most bow-shock
there is a clear correspondence with the compact source called IRAS16547S (Table ~\ref{table1}), which show polarization. 
This 1.2 mm continuum source additionally coincides well with the thermal free-free jet in its southern side. 
The polarization vectors are tracing also a bow-shock or outflow cavity structure with a similar orientation to the one traced by the H$^{13}$CO$^+$(3$-$2) emission. Perhaps they are part of the bow-shock.  
 As mentioned earlier, in this image we also include the Moment 1 map that traces the radial velocity field of the region.  There are a mixed of radial velocities, some molecular arcs are blueshifted, 
 and the others redshifted in both sides of the bipolar thermal free-free jet, likely because the multiple ejections.  The magnetic field vectors are associated with 
 blueshifted, and redshifted radial velocities in the inner part of the bi-conical envelope. There is no distinction between radial velocities.    
 
Figure~\ref{fig:fig4}, is similar to Figure~\ref{fig:fig3}, but now we present the emission from the HN$^{13}$C(3$-$2) that is also an optically thin line transition. In this Figure, we include  
the Moment 0 (integrated intensity), and Moment 1 (intensity-weighted velocity image) maps of the molecular emission overlaid with the polarization vectors that are tracing the magnetic 
fields in IRAS16547$-$4247, the continuum emission at 1.2 mm, and the thermal free-free jet, as before. To compute these moment maps 
we integrate over radial velocities from $-$49.7 km s$^{-1}$ to $-$11.3 km s$^{-1}$.  We integrate in smaller velocity window because the line-width for HN$^{13}$C(3$-$2) 
line is narrower as compared with the H$^{13}$CO$^+$(3$-$2) line. From these moments , the molecular thermal emission is tracing again the envelope with bi-conical outflow cavities observed as in the 
H$^{13}$CO$^+$(3$-$2) line, and the 1.2 mm continuum emission. However, the emission from the HN$^{13}$C(3$-$2) is much more compact, and is totally related to the 1.2 mm continuum emission.
From this spectral line, it is much clear to see the radial velocity field in the innermost part of IRAS16547$-$4247. There is a clear velocity gradient from southeast to northwest, with the blueshifted 
velocities toward the east, and the redshifted velocities localized in the northwest. We remark that this gradient was also noted in \citet{zap2015}, 
and was interpreted to be produced by the thermal (free-free) jet emerging from this object.  The large velocity gradient (about 20 km s$^{-1}$) could be associated with the outflow.  

In Figure~\ref{fig:fig5}, we present the Moment 8 and 1 maps (maximum value of the spectrum, and radial velocity fields, respectively) of the CS(5$-$4) molecular emission present in our spectral setup. 
Additionally, we also include in this Figure the continuum emission at 1.2 mm, and the thermal NS free-free thermal jet. To compute these moment maps we integrate 
over radial velocities from $-$64.4 km s$^{-1}$ to $+$9.64 km s$^{-1}$. The broader velocity range compared to the other two spectral lines is likely due to 
CS is tracing more clearly the outflows in the region. From this image, it is clear to see the different ejections from the bipolar thermal jet, and perhaps 
from other outflows located in the region.  \citet{Hig2015} proposed the existence of two more high-velocity CO outflows whose driving sources are located within 
the dust continuum peak. The orientation of these extra CO outflows do not seem to coincide with that of the wide-angle, large-scale, bipolar outflow detected 
with APEX \citep{gar2007} nor the thermal free-free jet \citep{rod2008}. These CO outflows could explain the different ejections observed in Figure~\ref{fig:fig5}.  A second possibility is that the 
NW-SE bipolar outflow has a strong precession as observed by  \citet{bro2003, rod2008}. From Figure ~\ref{fig:fig5}, the thermal free-free jet should precessing 
more than 150$^\circ$ (see the aperture angles) on the plane-of-sky. Maybe the massive binary system IRAS16547-Ea and IRAS16547-Eb is responsible of this large precession. 

Finally, a channel velocity map of the CS(5$-$4) molecular emission is presented in Figure~\ref{fig:fig6}.  This image reveals the complex outflow zone in IRAS16547$-$4247. 
This map reveals the structure and radial velocities of every bow-shock and outflow cavity present in  Figure~\ref{fig:fig5}.
 In particular, the bow-shocks and outflow cavities located toward south and north are blue/red-shifted, with not a clear difference in both sides.  
 This can be explained if the precession is occurring far from the plane-of-the-sky, with the ejections 
 having receding and approaching radial velocities toward the south and north sides. Precession in outflows from massive protostars seems is fairly common.    
 \citet{god2020,zap2013,maite2016} proposed this physical mechanism in W51 North, Cepheus A East, and G31.41+0.31, respectively, to explain the different powerful mass ejections.

\subsection{Magnetic Field Strength} \label{sec:bfields}

\begin{figure*}[t!]
\centering
\includegraphics[angle=-90,width=0.95\linewidth]{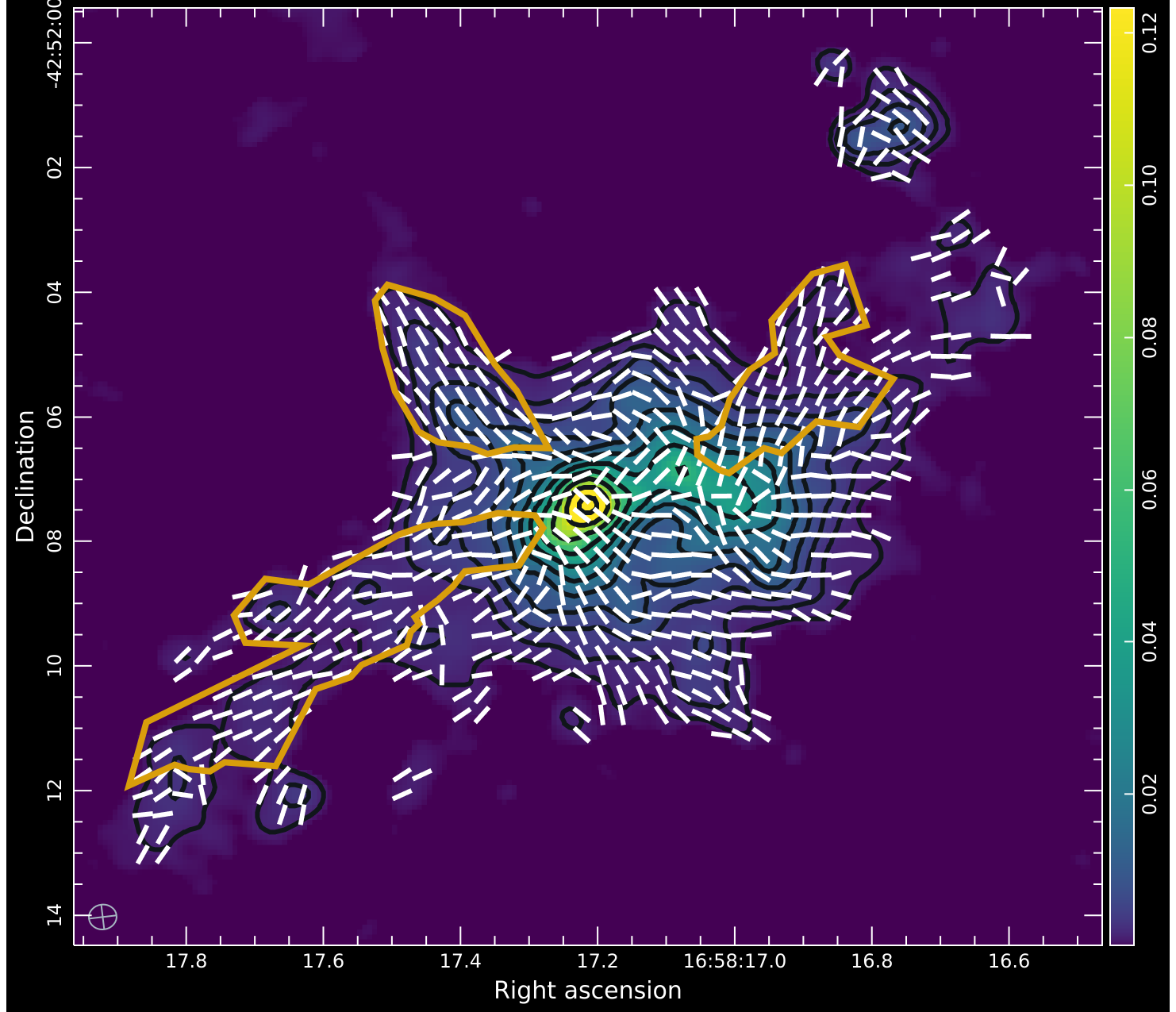}
\caption{Stokes I continuum image of toward the IRAS16547 region (contours and colour-scales), overlayed with magnetic field orientation sticks (white segments) 
and the three filament-like regions (yellow polygons) analyzed in Section \ref{sec:bfields} to derive an order of magnitude for the strength of the magnetic fields.
The synthesized beam (half-power contour) of the ALMA continuum image is shown in the bottom-left corner.  The scale bar in the right represents the peak flux of the ALMA continuum emission.  }
\label{fig:regions}
\end{figure*}

In this Section, we use the Davis-Chandrasekar-Fermi method  \citep[DCF;]{dav1951,cha1953} to estimate the magnetic field strength within some structures 
that are indicated in Figure \ref{fig:regions}. The overall morphology of the magnetic field toward IRAS16547 is complex, but we could observe elongated filament-like
 structures where the fields are uniform, whereas close to the main two protostellar centers, the fields are tangled into hourglass combined with spiral patterns (see Figure~\ref{fig:fig1}). 
 The filaments are identified as velocity coherent structures in at least two different molecular transtions (CS(5-4) and HN$^{13}$C(3-2)), but avoiding at the same time pixels where the magnetic field is very  twisted from the main direction (e.g., very close to the protostars) to preserve as much as possible the DCF assumptions. 
 From now on, we focus on the filament-like structures with fairly uniform fields that are also coherent in radial velocity, since a model to mimic all this complexity is beyond the scope of this paper 
 and will be addressed elsewhere. For these filament-like structures, the derivation of the magnetic field strength follows a standard pathway 
 \citep[e.g.,][]{Bel2019,San2021}, which relies on the expression:

$$B_{pos}\sim\xi\frac{\sigma_{los}}{\delta\psi}\sqrt{4\pi\rho}$$.

\begin{deluxetable*}{ccccccccccc}
\tablewidth{0pt}
\tablecolumns{11}
\tabletypesize{\scriptsize}
\tablecaption{Filament-like regions used to derive the magnetic field strength $B_{pos}$ from the HN$^{13}$C observations using the DCF method.}
\tablehead{
\colhead{ID} & \colhead{L$_{cyl}$\tablenotemark{*}} & \colhead{r$_{cyl}$\tablenotemark{*}} & \colhead{$T$} & \colhead{$M$} & \colhead{$n_{H2}$\tablenotemark{\dag}} & \colhead{$\sigma_{los}$} & \colhead{$\delta\Psi$} & \colhead{$B_{pos}$} & \colhead{$v_A$} & \colhead{E$_{turb}$/E$_{mag}$}\\
\colhead{} & \colhead{[$\arcsec$]} & \colhead{[$\arcsec$]} & \colhead{[$10^{-17}$ g~cm$^{-3}$]} & \colhead{[$10^{6}$ cm$^{-3}$]} & \colhead{[km s$^{-1}$]} & \colhead{[$\degr$]} & \colhead{[mG]} & \colhead{[km s$^{-1}$]} & \colhead{}  } 
\startdata
North & 7.4 & 0.40 & 16    & 15    & 1.89    & 0.26 & 15.3 & 2.3     & 0.7 & 0.2 \\
East  & 3.2 & 0.65 & 13-16 & 23-31 & 3.3-4.4 & 0.23 & 10.1 & 2.0-2.4 & 0.5 & 0.4 \\
West  & 4.4 & 0.45 & 13-16 & 20-27 & 3.8-5.1 & 0.21 &  8.8 & 5.2-6.1 & 1.1 & 0.1 \\
\enddata 
\tablenotetext{*}{L$_{cyl}$ and r$_{cyl}$ are the approximate length and radius of the filament-like structures.}
\tablenotetext{\dag}{The density is estimated under the assumption that the 3D-morphology of these regions is cylindrical, and using two different dust 
temperature values for the East and West filaments; hence, the two values for columns 4, 5, 6 and 9. For columns 8 and 11, the derived values using 
both temperatures coincide taking into account the uncertainties, and only one value is provided.}
\label{t:bpos}
\end{deluxetable*}

In this equation, $B_{pos}$ stands for the plane-of-the-sky component of the magnetic field strength, $\xi=1/2$ is a correction factor derived from
simulations of turbulent clouds \citep{2001eve, liu2021}, $\sigma_{los}$ is the velocity dispersion in the line of sight, $\delta\psi$ is the dispersion of the
magnetic field angles, and $\rho$ is the density. Let us make the caveat that there are more appropriate correction factors for cylinders. For instance, 
in the case of strong magnetic fields $\xi=0.22$ \citep{liu2021}. Since there is no \textit{a priori} information on the magnetic fields on this region, 
we use the classical value for $\xi$, but acknowledging that the field strengths could be a factor of two lower only due to geometry issues.  
The three selected regions appear as distinct dust filaments with a roughly uniform magnetic field orientation aligned along their major axes. 
The morphology of the polarization segments eases the usage of the DCF method, since we can assume a field with constant orientation. 
We have measurements of dust continuum, polarization angles, and the optically thin gas tracer HN$^{13}$C covering most of these regions. 
Each filament-like region comprises more than 40 independent Nyquist-sampled measurements, enough 
to secure a margin of error $<3\degr$ with a 95\% reliability.

From the equation 1 and the values presented in the that Section,  we derive the density ($\rho$) and the total mass from the dust continuum emission 
from these three regions shown in Figure \ref{fig:regions}. We assume that the dust temperature coincides with the gas temperature,
which we measure at the position with the maximum brightness of the filaments within the velocity cube. While most of the emission 
from the filaments is optically thin in all velocity channels ($\tau<0.2$), the peak emission of the three filaments is marginally optically thick. 
We use the optically thick maximum emission to estimate the gas temperature, whereas assume optically thin emission in the filaments 
to measure the density and mass. The resulting temperature ranges 13-16~K. Using these,  the total dust and gas mass enclosed in the filament-like
regions amounts 55-75 M$_\odot$. Assuming a cylindrical volume for each selected region, we find that the density varies 
between $0.9$ and $2.4\times10^{-16}$\,g~cm$^{-3}$, which correspond to number densities between 1.9 and 5.1$\times10^{7}$~cm$^{-3}$.

We extract the mean velocity dispersion from the optically thin tracer HN$^{13}$C after deconvolving with the channel width (0.28 km s$^{-1}$), 
and after subtracting in quadrature the thermal component broadening the line, $\sigma_{los}=\sqrt{\sigma^2_{observed}-\sigma^2_{thermal}}$. 
The thermal component of the lines is derived using $\sigma_{thermal}=\sqrt{(K_B T)/(\mu m_H)}$, where $K_B$ is the Boltzmann 
constant, $m_H$ is the atomic hydrogen weight, $T$ is the temperature of the gas, and $\mu=28.350$ is the weight of the HN$^{13}$C molecule. 
At 16~K, $\sigma_{thermal}$ takes the value of 0.07 km s$^{-1}$.

We estimate the magnetic field strength in three different regions extracted from the IRAS16547 field of view.
These regions were selected because the magnetic field orientations were quite uniform within them, which eases the usage of the DCF method.
We have measurements of dust continuum, polarization angles, and the optically thin gas tracer HN$^{13}$C covering the entirety of these regions, which
allows to estimate the magnetic field strength. The average effective radius of the three regions ($r_{eff}=\sqrt(A/\pi)$, where $A$ is the area) is $0\farcs68$,
or about 2000\,au at the adopted source distance. The smallest of these regions surrounds 1000\,au centered at the position of the main continuum sources;
we removed it from the average values reported below. This way, we can have an idea of the magnetic field strength at different distances from the source
center and at a range of densities.

Likewise, to derive the dispersion of the magnetic field angles ($\delta\psi$) we subtract in quadrature the mean angle uncertainty of the measurements ($\delta\psi_{obs}$) 
to the standard deviation of the polarization angles ($\sigma_{\psi}$) found for every specific region: $\delta\psi=\sqrt{\sigma^2_{\psi}-\delta\psi^2_{obs}}$. 
$\sigma_{\psi}$, the intrinsic angle dispersion, is estimated as the standard deviation of the polarization angles in each region, under the uniform magnetic field assumption. 
The mean value of $\delta\psi_{obs}$ (derived as $0.5~\sigma_{QU}/\sqrt{Q^2+U^2}$ in a pixel-by-pixel basis, with $\sigma_{QU}$ the noise of the Stokes Q and U images)
is $7\fdg6$, within the three regions considered.

The derived mean magnetic field strength in the regions goes from 2.0 to 6.1\,mG, and the Alfv\'en speed, given by $v_A=B/\sqrt{4\pi\rho}$, from 0.5 to 1.1 km s$^{-1}$.  
From here, we extracted the ratio between the turbulent and the magnetic  energies: E$_{turb}$/E$_{mag}$=$3\sigma_{los}^2/v_A$. We obtain values between 0.1 and 0.4, suggesting that the 
magnetic energy would be larger than turbulence in these filaments (\ref{t:bpos}). Given the high uncertainty dominated by the finite spectral resolution (which dominates the error 
propagation in the B field strength measurements, $\approx0.4-0.7$~mG), and the mass estimate (which adds at least another 25\% uncertainty), the values of these 
three quantities (B$_{pos}$, v$_A$, and E$_{turb}$/E$_{mag}$) have to be used with extreme caution.

\section{Summary}

We presented high angular resolution ($\sim$ 0.4$''$), and sensitive 1.2 mm dust polarization, and molecular line (HN$^{13}$C(3$-$2), CS(5$-$4), and H$^{13}$CO$^+$(3$-$2))
 Atacama Large Millimeter/Submillimeter Array observations of the rotating, and massive hot molecular core embedded in the high-mass star-forming region IRAS16547$-$4247.  
 The main conclusions of this study are as follows.  

\begin{itemize}

\item In the 1.2 mm continuum Stokes I emission, we reveal a dusty envelope surrounding the massive protostars,  IRAS16547-E and IRAS16547-W, with dimensions of 10,000 AU. 
This envelope has a bi-conical structure likely carved by the powerful thermal radio jet present in region.  We estimated a mass, and volumetric density for the bi-conical 
envelope of 220 M$_\odot$ and 2 $\times$ 10$^{8}$ cm$^{-3}$, respectively.   

\item We report the magnetic field orientations of the bi-conical envelope in IRAS16547$-$4247.  
The magnetic fields vectors follow very well the structure of the continuum emission or the bi-conical envelope. Some of these vectors seem 
to converge to IRAS 16547-E and IRAS 16547-W, the massive protostars. Some these polarization vectors likely are tracing the cavities of the outflows 
reported in the region. However,  some of them could also be tracing in-fall material, but more observations are need to confirm this possibility.  

\item The velocity fields revealed by the optically thin spectral lines (e.g. HN$^{13}$C) showed velocity gradients with a good correspondence with the magnetic fields. 
  
\item The CS(5$-$4) molecular line emission reveals multiple ejections with different orientations, some of them seem to follow the NW-SE radio thermal jet.
 All these ejections could be explained if the binary massive system in the middle of IRAS16547$-$4247 is precessing strongly, causing the motion 
 of the bipolar outflow.  

\item We derived a magnetic field strength in the regions goes from 2 to 6.1\,mG (the latter found associated to the central source position).

\end{itemize}

\begin{acknowledgments}
We thank the anonymous referee for providing a constructive report, which indeed improved the paper.
This paper makes use of the following ALMA data: ADS/JAO.ALMA\#2017.1.00101.S ALMA is a partnership of ESO (representing its member states), 
NSF (USA) and NINS (Japan), together with NRC (Canada), MOST and ASIAA (Taiwan), and KASI (Republic of Korea), in cooperation with the 
Republic of Chile. The Joint ALMA Observatory is operated by ESO, AUI/NRAO and NAOJ.  L.A.Z. acknowledges financial support from 
CONACyT-280775, UNAM-PAPIIT IN110618, and IN112323 grants, México. L.F.R. acknowledges the financial support of DGAPA (UNAM) IN105617,
IN101418, IN110618 and IN112417 and CONACyT 238631 and 280775-CF.
grant 263356. P.S. was partially supported by a Grant-in-Aid for Scientific Research (KAKENHI numbers JP22H01271 and JP24K17100) of the Japan 
Society for the Promotion of Science (JSPS). Y.C. was partially supported  by a Grant-in-Aid for Scientific Research (KAKENHI  number JP24K17103) of the JSPS.
M.T.B. acknowledges financial support through the INAF Large Grant {\it The role of MAGnetic fields in MAssive star formation}  (MAGMA). 
K.P. is a Royal Society University Research Fellow, supported by grant number URF\ R1\ 211322. MFL acknowledges support from the European Research 
Executive Agency HORIZON-MSCA-2021-SE-01 Research and Innovation programme under the Marie Skłodowska-Curie grant agreement number 101086388 
(LACEGAL). MFL also acknowledges the warmth and hospitality of the ICE-UB group of star formation.
\end{acknowledgments}

\facilities{ALMA}
\software{CASA \citep{cac2022}}

\bibliography{intro}{}
\bibliographystyle{aasjournal}

\end{document}